\documentclass[a4paper]{jpconf}
\usepackage{graphicx}
\usepackage{amsmath}
\usepackage{amssymb}
\usepackage{cite}

\DeclareMathOperator{\sech}{sech}

\begin{document}
\title{Corrections to Newton's law of gravitation - application to
hybrid Bloch brane}

\author{C A S Almeida, D F S Veras  and D M Dantas}

\address{Physics Department - Universidade Federal do Cear\'{a}, Fortaleza, CE, BRAZIL}

\ead{carlos@fisica.ufc.br, franklin@fisica.ufc.br, davi@fisica.ufc.br}

\begin{abstract}
We present in this work, the calculations of corrections in the Newton's law of gravitation
due to Kaluza-Klein gravitons in five-dimensional warped thick braneworld scenarios.
We consider here a recently proposed model, namely, the hybrid Bloch brane. This
model couples two scalar fields to gravity and is engendered from a domain wall-like
defect. Also, two other models the so-called asymmetric hybrid brane and compact brane
are considered. As a matter of fact, these models are obtained from deformations of the $\phi^4$ and sine-Gordon topological defects. Then, we constructed the branes upon such defects, and the corresponding corrections in Newton's law of gravitation are computed. In order to attain the mass spectrum and its corresponding eigenfunctions which are the essential quantities for computing the correction to the Newtonian potential, we develop a suitable numerical technique. 
\end{abstract}

\section{Introduction}
Phenomenological application to
braneworld models is a very interesting current subject. As can be found in detail in
 the introduction of reference \cite{ce-degenerate}, we have recent works in the braneworld branch explaining the process in the
meson $B$ decay  and in the neutrinos physics,
setting bounds into corrections to Coulomb's law and into
electrical conductivity. Selecting parameters to the
experimental data of the nucleon-nucleus total cross-section of various
chemical elements, and interesting applications to black holes and gravitational waves, are other examples.

In this view, one of most famous scenarios is the Bloch brane \cite{bloch-brane}. This smooth model consists in a thick brane generated by
two interacting scalar fields which solve several issues in the fields localisation present in thin
models \cite{rs, bajc}, especially in the context of gauge fields and fermion fields. The Bloch brane allows the existence of internal structures and a rich variety of scalar field deformations. So, there are many other models based on the initial proposal of the Bloch brane, such the degenerate Bloch brane \cite{ce-degenerate}  and recently the hybrid Bloch brane \cite{hybrid-bloch}. The term ``hybrid'' comes from the existence of an interval with compact structure \cite{compacton}. One of the scalar fields that compose the Bloch model is the well-known kink solution, which describes a wide variety of physics phenomena by connecting smoothly two minima. Unlike the kink where the energy density vanishes asymptotically, a compact
kink is a linear compact structure, with the energy density that vanishes outside a closed interval \cite{compacton}. Then, the Bloch brane proposed in the reference \cite{hybrid-bloch} is the so-called hybrid, because exhibits a usual thick brane inside an interval and a compact solution outside of this interval. We analyse in this work the gravity localization and a perspective to the application of corrections to Newton's law in the hybrid Bloch brane model.

In next section \ref{s-2}, we made a brief review of the usual Bloch brane of reference \cite{bloch-brane}, next we present the two versions of the hybrid Bloch brane of reference \cite{hybrid-bloch}. In the section \ref{s-3} we introduce the general way to confine gravity in a 5D warped model and comment how to obtain the expression of correction to gravity potential. Finally, in the conclusion, we point out the main points and results of this work.

\medskip
\section{The Hybrid Bloch brane}\label{s-2}
\medskip
\subsection{The usual Bloch brane}
The Bloch brane is built upon a 5D warped geometry with the metric \cite{bloch-brane}:
\begin{equation}\label{metric}
ds^2=g_{ab}dx^{a}dx^{b}=\e^{2 A(y)}\eta_{\mu\nu}dx^{\mu}dx^{\nu}-dy^2 \ , 
\end{equation}
where the indexes $a,b=0,1,2,3,4$, the $A(y)$ is the warp factor and the $\eta_{\mu\nu}$ is the 4D Minkowski metric  ($\mu,\nu=0,1,2,3$).

The action for this model is written in this form  \cite{bloch-brane}:
\begin{equation}\label{action}
I=\int d^4xdy \sqrt{\vert g \vert}\left(-\frac{R}{4}+\mathcal{L}_s\left(\phi, \chi\right)\right) \ ,
\end{equation}
where the $R$ represents the scalar curvature and the two scalar fields Lagrangian is given by:
\begin{equation}\label{lagrangian}
\mathcal{L}_s\left(\phi, \chi\right)=\frac{1}{2}\left(\partial_a\phi\partial^a\phi+\partial_a\chi\partial^a\chi+2 V\left(\phi, \chi\right)\right)
\end{equation}

For the Bloch brane, the scalar fields $\phi, \chi$ are only dependent on extra dimension coordinate $y$, so the equation of motions reads \cite{bloch-brane}:
\begin{gather}
\phi'+4A'\phi'=\frac{\partial V}{\partial \phi}, \nonumber\\ \chi'+4A'\chi'=\frac{\partial V}{\partial \chi}, \nonumber \\
A'^{2}=-\frac{1}{6}\left(\phi'^2+\chi'^2\right)-\frac{1}{3}V ,  \\ A''=-\frac{2}{3}\left(\phi'^2+\chi'^2\right), \label{eq-of-motion}
\end{gather}
the primes denote the derivative with relation to $y$ coordinate.

 The above equation of motion can be simplified to first order equations if the so-called superpotential $W$ is considered \cite{hybrid-bloch}:
\begin{equation}\label{superpotential}
V\left(\phi, \chi\right)=\frac{1}{2}\left[\left(\frac{\partial W}{\partial \phi}\right)^2+\left(\frac{\partial W}{\partial \chi}\right)^2-\frac{8}{3}W^2\right] \ .
\end{equation}

Thus, the first-order version of equations \eqref{eq-of-motion} with the superpotential \eqref{superpotential} becomes:
\begin{equation}\label{eq-of-motion-2}
\phi'=\frac{\partial W}{\partial \phi}, \ \chi'=\frac{\partial W}{\partial \chi}, \ A'=-\frac{2}{3} W \ .
\end{equation}

\begin{figure}[h]
\centering
\includegraphics[width=20pc]{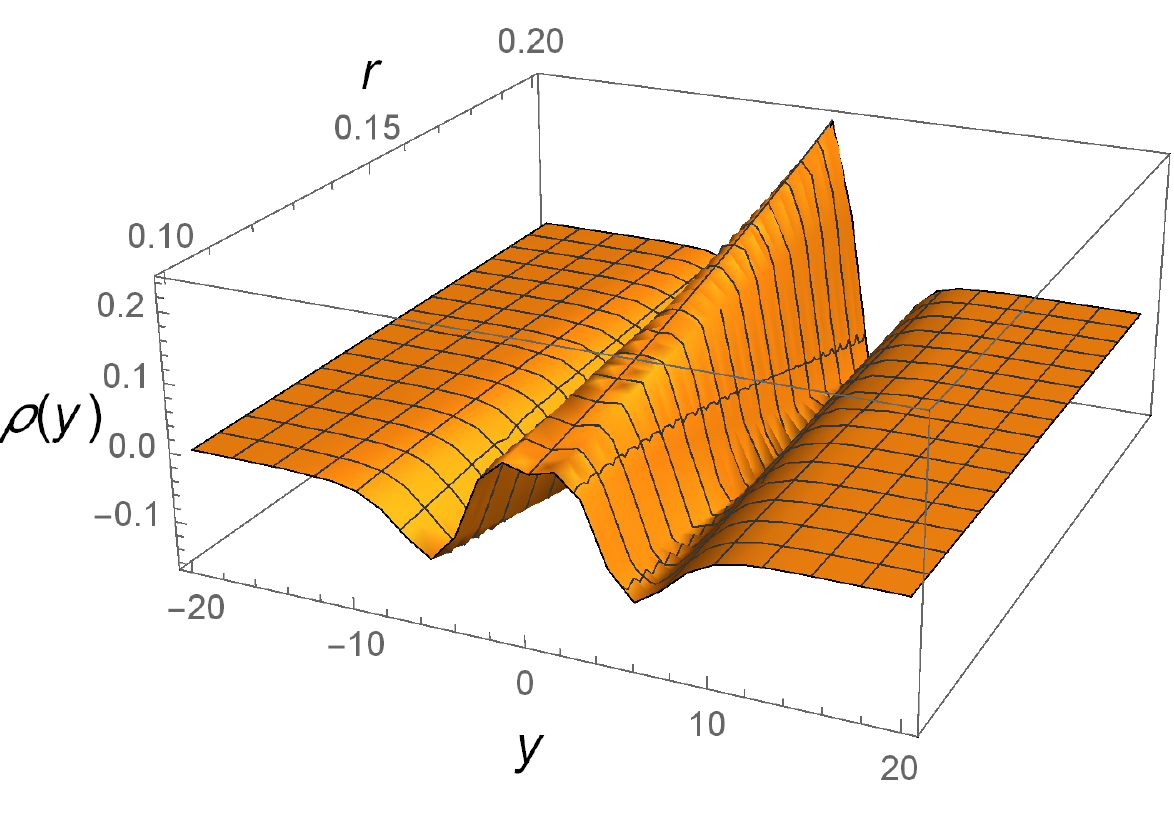}\hspace{2pc}\\
\begin{minipage}[b]{20pc}\caption{\label{energy-f} Energy density to usual Bloch brane of equation \eqref{energy}. The parameter $r \in(0.1,0.2)$ in this plot. We verify a split of energy density for small values of $r$.}
\end{minipage}
\end{figure}

Hence, the Ref. \cite{bloch-brane} defines the following superpotential to the Bloch Brane as:
\begin{equation}
W(\phi, \chi)=\phi-\frac{1}{3}\phi'^3-r \phi \chi^2 \ , \label{bloch-super}
\end{equation}
where the real parameter $r$, limited to $r \in (0,1)$, regulates the regimen of one field solution to the two fields solution. The choice of superpotential in equation \eqref{bloch-super} allows the analytical solution to warp factor and fields as follows \cite{bloch-brane, hybrid-bloch}:
\begin{gather}\label{solutions}
A(y)=\frac{1}{9r}\left[(1-3r)\tanh^2(2ry)-2 \ln \cosh(2ry)\right], \nonumber \\
\phi(y)=\tanh(2ry), \\ \chi(y)=\pm \sqrt{\left(\frac{1}{r}-1\right)}\sech(2ry)\nonumber \ .
\end{gather}
Note that the profile of $\phi$ field is a kink-like solution, while the $\chi$ field exhibits a lump-like profile. These solutions provide a thick brane scenario with localised energy given by\cite{bloch-brane}
\begin{gather}\label{energy}
\rho(y)=\e^{2 A(y)}\left[\frac{1}{2}\left(\frac{d\phi(y)}{dy}\right)^2+\frac{1}{2}\left(\frac{d\chi(y)}{dy}\right)^2+V\left(\phi(y),\chi(y)\right)\right] \ .
\end{gather}
The Bloch brane energy configuration can presents a split according to the $r$ parameter, as can be seen in figure \ref{energy-f}. For more details and physical applications to this Bloch brane model, we recommend the work \cite{ce-degenerate}.

\medskip

\subsection{Symmetric and asymetric hybrid Bloch brane}
Other choices to the superpotential $W(\phi,\chi)$ will be analysed in this section. The first case is the one that provides a symmetric profile with respect to an axis where the brane is placed. So, the  warp factor and the energy density are even function in this case. This is written by modifying the usual Bloch superpotential in equation \eqref{bloch-super} as follows\cite{hybrid-bloch}:
\begin{equation}\label{bloch-super-s}
W(\phi, \chi)=\phi-\frac{\phi^{2n+1}}{2n+1}-r\phi\chi^2\ ,
\end{equation}
where $n$ is a positive integer parameter that, for large values, split the brane in two intervals. We have two domains, one inside interval where the brane is non-compact, and other outside interval where the brane is compact. Due to this behaviour, this model is called Hybrid Bloch Brane \cite{hybrid-bloch}. Moreover, for $n=1$ the usual Bloch brane superpotential \eqref{bloch-super} is retrieved.  The reference \cite{hybrid-bloch} shows analytically that for $n\to\infty$ the model achieves, exactly, the thin brane profile in the outside interval due to compact behaviour. 

\begin{figure}[h]
\centering
\includegraphics[width=20pc]{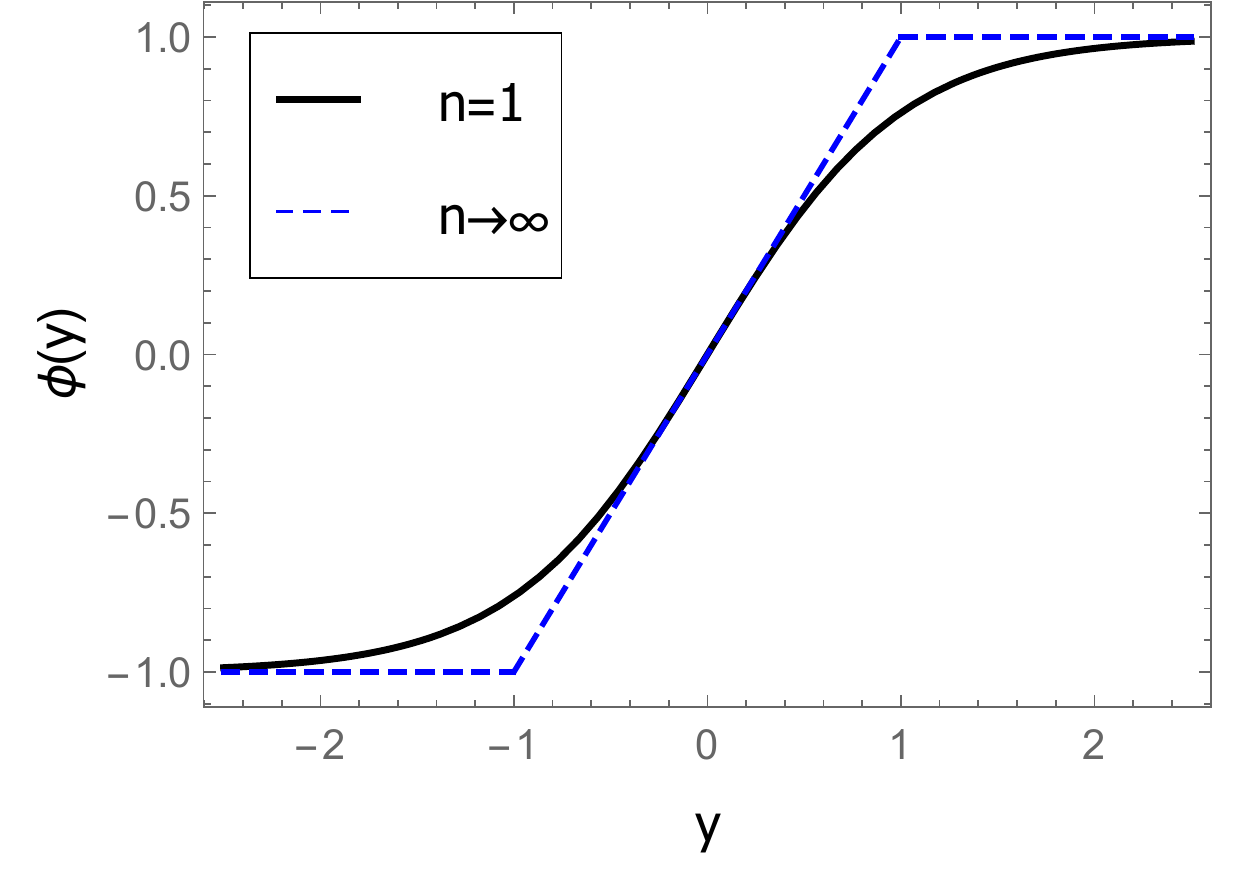}\hspace{2pc}\\
\begin{minipage}[b]{20pc}\caption{\label{kink} Differences between the kink and the compacton.}
\end{minipage}
\end{figure}

In order to illustrate the differences between the usual structure and the compact structure we compare the $\phi$ field in the kink-like profile $(n=1)$ to the half-compact profile $(n\to\infty)$ in figure \ref{kink}. Note in the Fig. \ref{kink} that while the kink (black full line) achieves the points $\phi(y\to\pm\infty)=\pm 1$ the kink compacton (blue dashed line) achieves  $\phi=\pm 1$ for $\lvert y \lvert>1$.

Another choice to the superpotential is the one that exhibits an asymmetric profile to the brane, obtained as:
\begin{equation}\label{bloch-super-a}
W(\phi, \chi)=\phi-\frac{\phi^{2}}{2}+ \frac{\phi^{p+1}}{p+1}-\frac{\phi^{p+2}}{p+2}-r\phi\chi^2\ .
\end{equation}
Here the $p$ parameter is an odd integer that also allows the existence of a compact interval to the brane, and also reproduces the usual Bloch brane superpotential  \eqref{bloch-super} when $p=1$. It is worthwhile to mention the important feature of asymmetric models in the massive modes of gravity. In reference \cite{coni} we verify that in these models the resonant state provides a stronger contribution to the correction in the Newtonian potential.

\section{Gravity localisation and the correction to Newton's Law}\label{s-3}

In order to obtain the correction to Newton's law on the hybrid Bloch brane scenario, we need to perform the gravity localisation by introducing a small fluctuation $h_{\mu, \nu}(\vec{x},y)$ in the metric \eqref{metric} in the form \cite{diego-1}
\begin{equation}\label{metric-2}
ds^2=\e^{2 A(y)}\left(\eta_{\mu \nu}+h_{\mu \nu}\right)dx^{\mu}dx^{\nu}-dy^2 \ .
\end{equation}
Now assuming the  transverse-traceless gauge and imposing that $h_{5N}=0$ \cite{diego-1}, we get the following equation of motion:
\begin{equation}\label{eqm-2}
h''_{\mu \nu} +4A'h'_{\mu \nu}=\e^{-2A} \square h_{\mu \nu} \ ,
\end{equation}
where the $\square$ is the 4D d'Alembertian. Setting the Kaluza-Klein (KK) decomposition \cite{diego-2}:
\begin{equation}\label{eqm-2b}
h_{\mu \nu}(\vec{x},y)=\sum_{k=0}^{\infty} \tilde{\eta}_{\mu \nu}^{(k)}\varphi_{k}(y) \ ,
\end{equation}
with the 4D portion
\begin{equation}\label{eqm-2bc}
\eta_{\mu \nu}\partial^{\mu}\partial^{\nu}\tilde{\eta}_{\mu \nu}^{(k)}=-m^2_k\tilde{\eta}_{\mu \nu}^{(k)} \ ,
\end{equation}
being $m_k$ the gravitational KK mass, we have the equation of motion in the form
\begin{equation}\label{eqm-3}
\varphi''_{k}(y) +4A'\varphi'_k(y)=-m_k^2\varphi_k(y) \ .
\end{equation}
Now, we use a conformal transformation in order to transform the equation \eqref{eqm-3} into a Schr\"{o}dinger-like equation. Now we use the independent variable transformation $y\to z$ \cite{diego-1}
\begin{equation}\label{vare}
\frac{dz}{dy}=\e^{-A(y)} \to z(y)=\int_{y}{\e^{-A(\tilde{y})}d\tilde{y}} \ .
\end{equation}
We choose that $z(y=0)=0$ in order to cancel the integration constant. Moreover, the dependent variable transformation is requested $\varphi\to \psi$\cite{diego-1}
\begin{equation}\label{vare2}
\varphi_k=\e^{-\frac{3}{2}A}\psi_k \ ,
\end{equation}
in order to get the following Schr\"{o}dinger-like equation \cite{diego-2}
\begin{equation}\label{eqm-4}
-\ddot{\psi}_{k}(y) +U(z)\psi_k(z)=m_k^2\psi_k(z) \ ,
\end{equation}
where the dot denotes derivative with relation to $z$ coordinate. The  analogue quantum potential has the form \cite{diego-1}
\begin{equation}
U(z)=\frac{3}{2}\ddot{A}(z)+\frac{9}{4}\dot{A}^2(z) \ .
\end{equation}

The general equation \eqref{eqm-4} give us all the massive (and massless) eigenstates for a specific model. The zero mode $(m=0)$ needs to be confined (normalized and holding all boundary conditions) in order to reproduce the usual 4D gravity at the brane. Moreover, none massive modes  can be normalized, therefore there is not a massive KK state bounded. However, some model can exposes a massive quasi-localised state that have high probability amplitude close to region where the brane is placed. These quasi-localised states are namely resonances. 

Independently of the existence of resonances, the massive KK modes perform a slight Yukawa-like correction to the Newton's potential in the form \cite{csaki, coni}:
\begin{equation}\label{newton}
V(x)=G_N\frac{m_1 m_2}{r}\left[1+\Delta(\vec{x})\right] \ ,
\end{equation}
where $x=\lvert \vec{x} \lvert$ is a distance in our 4D world, the $G_N$ is the gravitational constant, the $m_i$ are point-like source masses that live in the 4D space and the correction $\Delta(\vec{x})$ is obtained as
\begin{equation}\label{delta}
\Delta(x)=\sum_{k>0}^{\infty}\frac{\e^{-m_kx}}{x}\left.\left(\frac{\hat{\psi}_k(z)}{\hat{\psi}_0(z)}\right)\right\lvert_{z=z^*} \ ,
\end{equation}
where $\hat{\psi}_k$ represents the normalized massive eigenfunction and $\hat{\psi}_0$ the normalized zero mode (for $k=0$ this  quantity gives exactly $1$, the usual gravity, present in the equation \eqref{newton}). The $z^*$ is the point in the $z$ coordinate where the energy density is maximum. Usually, for symmetric non-splitted models this point match with the origin $z^*=0$.

The expressions of correction $\Delta(x)$ can be analytically obtained in some thin model, as the Randall-Sundrum (RS) model \cite{rs} and the Gherghetta-Shaposhnikov (GS) model \cite{gs}. In the RS model $\Delta(x) \approx (c x)^{-2}$ \cite{rs}, while in the GS model $\Delta(x) \approx (c x)^{-3}$ \cite{gs}, with $c$ as a parameter associated to the scalar curvature of the model.  In another models, this correction is dependent on more parameters, as the case of Resolved Conifold (RC) presented in reference \cite{coni1}. The RC case is phenomenologically more interesting since that other parameters can be combined in order to satisfy simultaneously the hierarchy problem, the correction of Newton's Law and other issues \cite{coni}.

\begin{figure}[h]
\centering
\includegraphics[width=20pc]{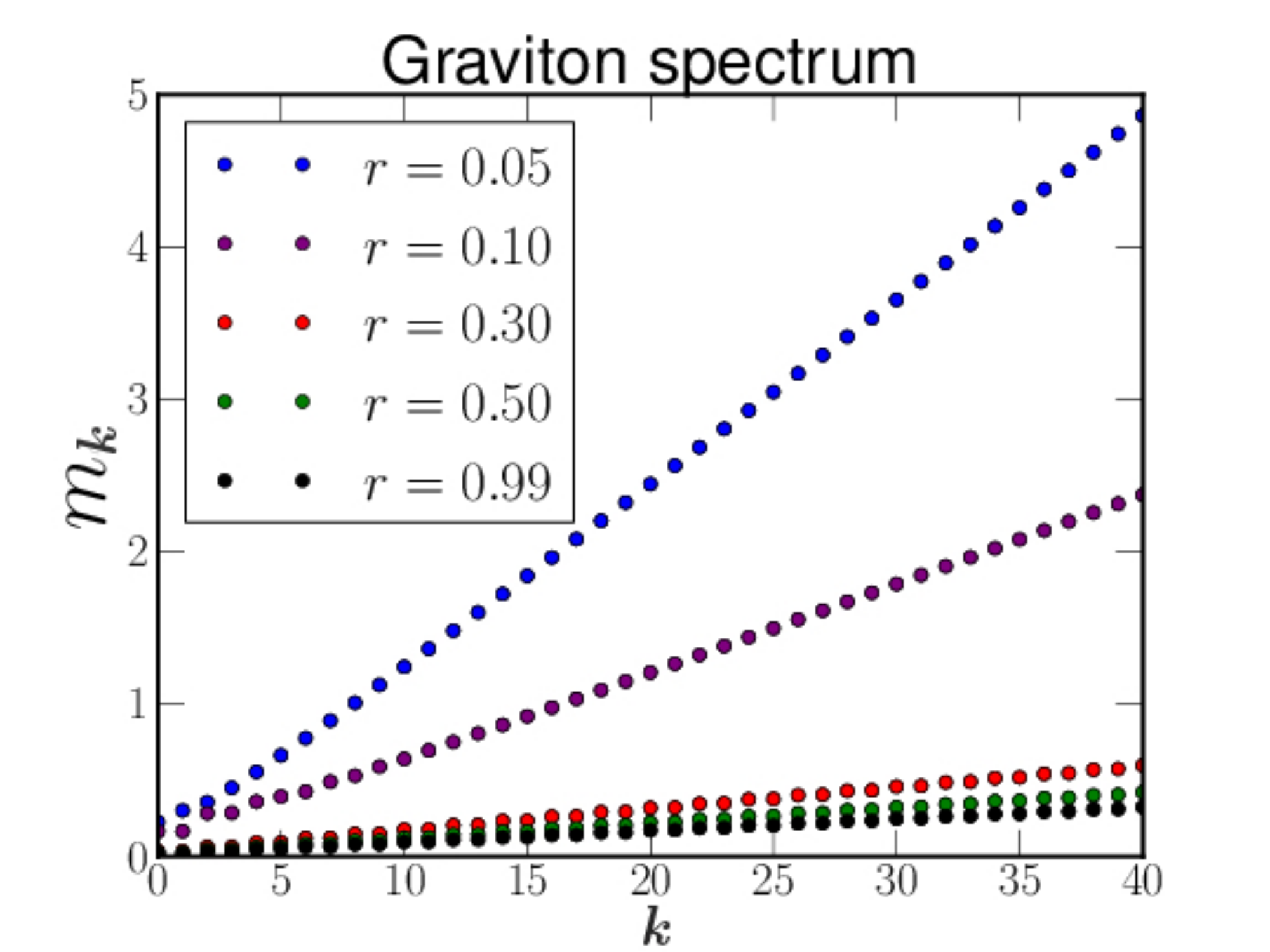}\hspace{2pc}\\
\begin{minipage}[b]{20pc}\caption{\label{massespctrum} Mass spectrum of the Bloch brane.}
\end{minipage}
\end{figure}

In this work, we expose an important result to achieve, namely, the correction of Newton's Law in the Bloch brane scenarios. The mass spectrum presented in figure \ref{massespctrum} is fundamental to set the discrete suitable mass value to eigenfunction of the Sch\"{o}dinger-like equation \eqref{eqm-4}. From figure \ref{massespctrum} we note that the mass spectrum is approximately linear and decreases for large $r$ parameter. Similar results are also found in reference \cite{diego-1}. After get the numerical results for the eigenfunction and applied it to equation \eqref{delta}, in the hybrid Bloch brane case, we can set experimental bound in the same way that performed in reference \cite{coni}. So, we can adjust the models parameters in order to obtain suitable results endorsed by the Newton's Law.


\medskip
\section{Conclusions}

In this work, we review the Bloch brane model and the recent version of the hybrid Bloch brane. We perform a general method to confine the gravity in a 5D warped model and present the expression to the correction of Newton's Law. Finally, we show the discrete mass spectrum to the Bloch Brane, which has some similarities to hybrid brane model as can be seen in Ref. \cite{diego-1}. For the sake of completeness as further steps, we intend to finish the numerical evaluation of massive eigenfunctions at the $z^*$ point in both symmetric and asymmetric hybrid Bloch brane in order to get the proper plot of gravity modification in these scenarios. We also intend to set experimental bounds in these corrections as presented in the work of Ref. \cite{coni}.

\ack{The authors thank the Coordena\c{c}\~{a}o de Aperfei\c{c}oamento de Pessoal de N\'{i}vel Superior (CAPES), the Conselho Nacional de Desenvolvimento Cient\'{i}fico e Tecnol\'{o}gico (CNPQ), and Funda\c{c}\~{a}o Cearense de Apoio ao Desenvolvimento Cient\'{i}fico e Tecnol\'{o}gico (FUNCAP) for financial support.}

\end{document}